\documentclass[conference]{IEEEtran}
\IEEEoverridecommandlockouts
\usepackage[braket, qm]{qcircuit}
\usepackage{graphicx}
\usepackage{physics}
\usepackage{amsmath,amssymb,amsfonts}
\usepackage{algorithmic}
\usepackage{graphicx}
\usepackage{textcomp}
\usepackage{xcolor}
\usepackage{subcaption}
\usepackage[moderate,tracking=normal]{savetrees}
\usepackage{amsmath}
 \usepackage[hidelinks]{hyperref}
\usepackage{pgffor}
\usepackage{adjustbox}

\usepackage[numbers,compress]{natbib}

\definecolor{quantumweek-1}{RGB}{182,214,186}
\definecolor{quantumweek-2}{RGB}{214,205,182}
\definecolor{quantumweek-3}{RGB}{182,190,214}
\definecolor{quantumweek-4}{RGB}{214,182,192}
\definecolor{quantumweek-5}{RGB}{81,93,130}
\definecolor{quantumweek-6}{RGB}{39,87,45}

\def\BibTeX{{\rm B\kern-.05em{\sc i\kern-.025em b}\kern-.08em
    T\kern-.1667em\lower.7ex\hbox{E}\kern-.125emX}}
\begin{document}

\title{Analyzing Common Electronic Structure Theory Algorithms for Distributed Quantum Computing\\
\thanks{We would like to acknowledge the Government of Canada’s New Frontiers in Research Fund (NFRF), for grant NFRFE-2022-00226, and the Quantum Software Consortium (QSC), financed under grant \#ALLRP587590-23 from the National Sciences and Engineering Research Council of Canada (NSERC) Alliance Consortia Quantum Grants.
}
}

\author{\IEEEauthorblockN{Grier M. Jones\IEEEauthorrefmark{1}\IEEEauthorrefmark{2}\IEEEauthorrefmark{3},Hans-Arno Jacobsen\IEEEauthorrefmark{2}}
\IEEEauthorblockA{
Department of Chemical and Physical Sciences\IEEEauthorrefmark{1}\\
The Edward S. Rogers Sr. Department of Electrical and Computer Engineering\IEEEauthorrefmark{2}\\
University of Toronto\\
Toronto, Ontario, Canada\\
Email: grier.jones@utoronto.ca\IEEEauthorrefmark{3}}

}

\IEEEoverridecommandlockouts 

\maketitle

\begin{abstract}
To move towards the utility era of quantum computing, many corporations have posed distributed quantum computing (DQC) as a framework for scaling the current generation of devices for practical applications.
One of these applications is quantum chemistry, also known as electronic structure theory, which has been poised as a ``killer application'' of quantum computing, 
To this end, we analyze five electronic structure methods, found in common packages such as Tequila and ffsim, which can be easily interfaced with the Qiskit Circuit Cutting addon.
Herein, we provide insights into cutting these algorithms using local operations (LO) to determine their aptitude for distribution.
The key findings of our work are that many of these algorithms cannot be efficiently parallelized using LO, and new methods must be developed to apply electronic structure theory within a DQC framework. 

\end{abstract}


\section{Introduction}
Recently, quantum chemistry has emerged as a leading candidate for demonstrating quantum advantage, driving efforts to map electronic structure problems onto quantum devices~\cite{aspuru-guzik_simulated_2005,cao_quantum_2019,mcardle_quantum_2020}.
To this end, methods such as the quantum phase estimation (QPE) algorithm~\cite{abrams_simulation_1997,abrams_quantum_1999,aspuru-guzik_simulated_2005,lanyon_towards_2010,whitfield_simulation_2011,aspuru-guzik_photonic_2012} have been shown to offer exponential speedups over classical methods.
Despite promising speedups, QPE requires long coherence times, while the current generation of quantum processing units (QPUs) are often too noisy for practical applications.
Alternatively, methods based on the variational principle, such as the variational quantum eigensolver (VQE)~\cite{peruzzo_variational_2014,cerezo_variational_2021,mcclean_theory_2016,bharti_noisy_2022}, have been proposed as a quantum-classical hybrid approach, capable of running on noisy, near-term quantum devices.
While the number of qubits limits the applicability of VQE on current devices, alternatives based on distributed quantum computing (DQC) have been proposed~\cite{khait2023variational}.

DQC offers promising alternatives to traditional quantum algorithms, where quantum algorithms can be distributed among QPUs with shallow circuits, few qubits, and reduced noise to achieve computational speedups~\cite{barral_review_2024,denchev_distributed_2008}.
While chemistry has benefited from access to classical distributed computing platforms, where parallelism can be exploited to accelerate electronic structure theory calculations using central processing units (CPUs)~\cite{lotrich_parallel_2008}, graphical processing units (GPUs)~\cite{gotz_chapter_2010}, or by cloud computing~\cite{raucci_interactive_2023}, examples of DQC within chemistry are lacking\cite{jones2024distributed}.
Despite the lack of development in the field of DQC for chemistry, DQC offers an attractive approach for solving electronic structure theory problems~\cite{reiher2017elucidating}.
To this end, we explore the efficiency of distributing common quantum chemistry algorithms, such as the unitary coupled-cluster singles and doubles (UCCSD)~\cite{kutzelnigg1991error,bartlett1989alternative,hoffmann1988unitary,hoffmann1988unitary,bartlett1988expectation}, unitary pair coupled-cluster double excitations (UpCCD)~\cite{lee_quantum_2022}, unitary pair coupled-cluster with generalized singles and doubles product wave functions (UpCCGSD)~\cite{lee_quantum_2022}, separable pair approximation with generalized singles (SPA+GS)~\cite{kottmann2022optimized}, and local unitary Jastrow (LUCJ)~\cite{motta2023bridging} ans\"{a}tze, using circuit cutting techniques~\cite{piveteau2023circuit}.
The key consideration for choosing these ans\"{a}tze is based on using packages, such as Tequila~\cite{kottmann2021tequila} and ffsim~\cite{ffsim}, that can easily be interfaced with Qiskit~\cite{qiskit2024} and the Qiskit Circuit Cutting addon~\cite{qiskit-addon-cutting}.

\section{Theoretical Framework}

\subsection{Molecular Electronic Structure Theory}
\begin{figure}[!ht]
    \centering
    \includegraphics[width=0.3\linewidth]{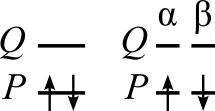}
    \caption{An example of spatial-orbitals (left), denoted as $P$ and $Q$, and spin-orbitals (right), denoted as $P\alpha$, $P\beta$, $Q\alpha$, and $P\beta$, for H$_{2}$ using a minimal basis. In this example, $P$ corresponds to the bonding ($\sigma$) and $Q$ corresponds to the anti-bonding ($\sigma^{\ast}$) orbitals shared between the two 1s orbitals of each hydrogen atom.}
    \label{fig:H2example}
\end{figure}

Molecular electronic structure theory is a subdiscipline of computational chemistry that focuses on solving the Schr\"{o}dinger equation to understand chemical phenomena~\cite{helgaker2013molecular}.
The Schr\"{o}dinger equation is defined as
\begin{equation}
    \hat{H}\Psi=E\Psi
    \label{eq:schrodingerEq}
\end{equation}
where $\Psi$ is a wave function describing the state of the system, $\hat{H}$ is the molecular Hamiltonian, and $E$ is the energy of the system, which is an eigenvalue of the Hamiltonian.
The molecular electronic Hamiltonian is defined using a notation that is spin-free, non-relativistic, and absent of an external field as
\begin{equation}
    \hat{H} = \sum_{PQ} h_{PQ}\hat{a}^{\dagger}_{P}\hat{a}_{Q} + \frac{1}{2}\sum_{PQRS} g_{PQRS} \hat{a}^{\dagger}_{P}\hat{a}^{\dagger}_{R}\hat{a}_{S}\hat{a}_{Q} + h_{nuc}
    \label{eq:mol_Ham}
\end{equation}
where $P,Q,R,S$ denote general spatial-orbitals (an example of spatial orbitals is shown in Fig. \ref{fig:H2example}), $\hat{a}_{Q}$ and $\hat{a}_{S}$ denote annihilation operators, that remove an electron from orbitals $Q$ and $S$, $\hat{a}^{\dagger}_{P}$ and $\hat{a}^{\dagger}_{R}$ denote creation operators that add an electron to orbitals $P$ and $R$.
The remaining terms, include the one-electron integral ($h_{PQ}$), two-electron integrals ($g_{PQRS}$), and nuclear-repulsion term ($h_{nuc}$).

In general, electronic wave functions can be represented in two forms, the first is referred to as a Slater determinant, defined as 
\begin{equation}
    \ket{\Psi} = \frac{1}{\sqrt{N!}}
    \begin{vmatrix}
    \phi_{P_{1}} (\mathbf{x}_{1}) & \phi_{P_{2}} (\mathbf{x}_{1}) & \cdots & \phi_{P_{N}} (\mathbf{x}_{1}) \\
    \phi_{P_{1}} (\mathbf{x}_{2}) & \phi_{P_{2}} (\mathbf{x}_{2}) & \cdots & \phi_{P_{N}} (\mathbf{x}_{2}) \\
    \vdots & \vdots & \ddots & \vdots \\
    \phi_{P_{1}} (\mathbf{x}_{N}) & \phi_{P_{2}} (\mathbf{x}_{N}) & \cdots & \phi_{P_{N}} (\mathbf{x}_{N}) 
    \end{vmatrix} \\
    \label{eq:slaterdet}
\end{equation}
where $\{ \phi_{P} (\mathbf{x})\}$ are a set of $M$ orthonormal spin-orbitals (an example of which are displayed in Fig. \ref{fig:H2example}), where the coordinates $\mathbf{x}$ can be partitioned into spatial ($\mathbf{r}$) and spin ($\sigma$) coordinates of an electron.
And the second representation, which uses the notation from second quantization, where the wave function is represented as an occupation-number vector,

\begin{equation}
    \ket{\Psi} = \ket{k_{1}, k_{2},\ldots,k_{M}}
    \label{eq:occnumvec}
\end{equation}

\noindent where $k_{p} = 1$ if $\phi_{P}$ is occupied and $k_{p} = 0$ if $\phi_{P}$ is unoccupied.

Often, the starting point for many electronic structure theory calculations is Hartree-Fock (HF).
Due to the mean-field nature of HF, a fraction of the Coulombic interactions between electrons, referred to as electron correlation, is neglected.
The energy associated with the missing electron correlation, or correlation energy, is defined as
\begin{equation}
    E_{\text{corr}} = E_{\text{exact}} - E_{\text{HF}},
    \label{eq:electron_corr}
\end{equation}
where, within a given one-particle basis set, the exact energy can be determined using the full configuration interaction (FCI) method.
The FCI wave function takes the following form
\begin{equation}
    \ket{\Psi_{\text{FCI}}} = \left( 1 + \sum_{AI} C^{A}_{I}\hat{a}^{\dagger}_{A}\hat{a}_{I} + \sum_{\substack{A>B\\I>J}} C^{AB}_{IJ}\hat{a}^{\dagger}_{A}\hat{a}^{\dagger}_{B}\hat{a}_{I}\hat{a}_{J} + \cdots \right) \ket{\Psi_{\text{HF}}}
    \label{eq:FCI_wfn}
\end{equation}
where electrons in the HF wave function ($\ket{\Psi_{\text{HF}}}$) are promoted from occupied orbitals $I,J, \ldots$ to unoccupied orbitals $A,B,\ldots$ with a configuration coefficient, defined as $C^{AB\cdots}_{IJ\cdots}$.
Due to the intractable nature of the problem as system size increases, the FCI expansion is often truncated to only include single (S), double (D), triple (T), and higher-order excitations on top of an HF wave function.
While truncated variants of FCI, such as configuration interaction singles and doubles (CISD), offer computationally tractable alternatives, these methods suffer from convergence issues towards the FCI limit and issues related to \textit{size-extensivity}.
For a more in-depth explanation of electron correlation, HF, and FCI, refer to Helgaker \textit{et al.}~\cite{helgaker2013molecular} for further information.

One family of methods that offers an attractive alternative to truncated CI variants is coupled-cluster (CC) theory~\cite{bartlett2007coupled,cizek1980coupled,vcivzek1966correlation}.
Within this family of methods, coupled-cluster singles and doubles with perturbative triples (CCSD(T)), is considered the \textit{gold standard} of quantum chemistry, due to the tradeoffs between accuracy and computational cost~\cite{helgaker2013molecular,townsend2019post}.
Unlike the CI wave function, the CC wave function takes an exponential form that is inherently separable, defined as

\begin{align}
    \begin{split}
        \ket{\Psi_{\text{CC}}} &= \exp\left(\sum_{AI} t^{A}_{I} \hat{a}^{\dagger}_{A}\hat{a}_{I} + \sum_{\substack{A>B\\I>J}} t^{AB}_{IJ} \hat{a}^{\dagger}_{A}\hat{a}^{\dagger}_{B}\hat{a}_{I}\hat{a}_{J} + \cdots  \right) \ket{\Psi_{\text{HF}}} \\
        &= e^{\hat{T}_{1}+\hat{T}_{2}+\cdots}\ket{\Psi_{\text{HF}}} =  e^{\hat{T}}\ket{\Psi_{\text{HF}}} \\
    \end{split}
\end{align}

\noindent where  $\hat{T}$ is a cluster operator, taking $\ket{\Psi_{\text{HF}}}$ to a state composed of a linear combination of excited electronic configurations, with amplitudes defined as $t^{AB\cdots}_{IJ\cdots}$.



One drawback of traditional CC is that it is not solved using a variational method, which limits its applicability on quantum computers. 
Alternatively, an extension of CC that is commonly applied and used as a reference point for other electronic structure methods in quantum computing is unitary coupled-cluster (UCC)~\cite{anand2022quantum,kutzelnigg1991error,bartlett1989alternative,hoffmann1988unitary,hoffmann1988unitary,bartlett1988expectation}.
The key difference between traditional CC and UCC is the inclusion of the de-excitation operator, $\Hat{T}^{\dagger}$, within the exponential ansatz, where the wave function is defined as
\begin{equation}
    \ket{\Psi} = e^{\Hat{T}-\Hat{T}^{\dagger}} \ket{\Phi_{HF}}.
    \label{eq:UCC_wfn}
\end{equation}
In this study, we examine the unitary coupled-cluster singles and doubles (UCCSD) method, where the cluster operator is truncated to include single and double excitations, defined as $\hat{T} = \hat{T}_{1} + \hat{T}_{2}$, on the HF wave function.



The next method we examine combines UCC with pair-coupled-cluster double excitation (pCCD) operators, known as the unitary pair coupled-cluster doubles (UpCCD) method~\cite{lee2018generalized}.
This method relies on a pair cluster operator, defined as 
\begin{equation}
    \hat{T} = \sum_{IA} t^{A_{\alpha} A_{\beta}}_{I_{\alpha} I_{\beta}} \hat{a}^{\dagger}_{A_{\alpha}}\hat{a}^{\dagger}_{A_{\beta}}\hat{a}_{I_{\beta}}\hat{a}_{I_{\alpha}}
    \label{eq:pairclusterop}
\end{equation}
which only includes two-electron excitations from the same occupied orbital $I$ to the same virtual orbital $A$.
Another variant of the cluster operator is the generalized cluster operator~\cite{nooijen2000brueckner}, 
\begin{equation}
    \hat{T} = \hat{T}_{1} + \hat{T}_{2} = \frac{1}{2} \sum_{pq} t^{q}_{p} \hat{a}^{\dagger}_{q}\hat{a}_{p} + \frac{1}{4} \sum_{pqrs} t^{rs}_{pq} \hat{a}^{\dagger}_{r}\hat{a}^{\dagger}_{s}\hat{a}_{q}\hat{a}_{p}.
    \label{eq:generalclusterop}
\end{equation}

\noindent which run over a general set of spin-orbitals, $p, q, r, s$.
Another variant of the UpCC examined in this study is, the unitary pair coupled-cluster with generalized singles and doubles (UpCCGSD) method, which incorporates both Eqs. \ref{eq:pairclusterop} and \ref{eq:generalclusterop} into one cluster operator.

The next method is the separable-pair approximation plus generalized singles (SPA+GS).
The SPA method was created as a generalization of the UpCC(G)SD ansatz using pair-natural-orbitals (PNO-UpCC(G)SD and incorporates the hard-core Boson model into the ansatz~\cite{kottmann2022optimized}.
The last method, examined is the local unitary cluster Jastrow (LUCJ) ansatz, proposed by Motta \textit{et al.} in 2023~\cite{motta2023bridging}, where the wave function takes the following form
\begin{equation}
    \ket{\Psi} = \prod_{\mu=1}^{L} e^{\Hat{K}_{\mu}}e^{i\Hat{J}_{\mu}}e^{-\Hat{K}_{\mu}} \ket{\Phi_{HF}}
    \label{eq:LUCJ}
\end{equation}
and the exponential cluster-Jastrow operator is defined as $e^{\hat{T}^{\text{CJ}}} = e^{-\Hat{K}}e^{\Hat{J}}e^{\Hat{K}}$~\cite{anand2022quantum}.







\subsection{Circuit Cutting}
To distribute the aforementioned algorithms across multiple quantum processing units (QPUs), circuit cutting using \textit{quasiprobability simulation} can be used.
Quasiprobability simulation involves replacing two-qubit (nonlocal) gates probabilistically with single-qubit (local) gates, where the expectation values from the measurement of the original circuit containing nonlocal gates are obtained by sampling the outcomes from the circuits with local operations.
Despite the number of nonlocal gate cuts incurring a cost, or sampling overhead, that scales exponentially with the number of gate cuts, this method can allow the implementation of quantum circuits that cannot be realized on current hardware due to the depth or width of the circuit examined~\cite{piveteau2023circuit}.

To realize quasiprobability simulation, three different communication protocols can be used:

\begin{enumerate}
    \item \textbf{Local operations (LO)}: which decomposes a two-qubit gate across devices using a tensor decomposition, e.g., the product form of the two-qubit gate, $A \bigotimes B$, is decomposed such that $A$ is run on one device and $B$ on another
    \item \textbf{Local operations and one-way classical communication ($\text{LO}\overrightarrow{\text{CC}}$)}: where two computers can perform LO, along with classical one-way communications from one device to another
    \item \textbf{Local operations and classical communication (LOCC)}: where two computers can perform LO, along with classical bidirectional communications between devices
\end{enumerate}
While LOCC requires two different devices to perform the bidirectional classical communication, LO and $\text{LO}\overrightarrow{\text{CC}}$ do not require separate QPUs and can be performed sequentially on the same QPU.

Nonlocal gates can be simulated with a unitary channel $\mathcal{U}$ using quasiprobability decomposition of the channel, i.e., a linear combination of quantum channels, defined as
\begin{equation}
    \mathcal{U} = \sum_i a_i \mathcal{F}_{i},
    \label{eq:QPD}
\end{equation}
where $\mathcal{F}_{i}$ are operations realizable on the hardware, depending on the communication protocol, such as $\mathcal{F}_{i}\in S=\{\text{LO},  \text{LO}\overrightarrow{\text{CC}}, \text{LOCC}\}$, and the coefficients $a_i$.
Note that $a_i$ are real numbers, which can be positive or negative, which is why it is referred to as a quasiprobability.
Throughout the process the gate $\mathcal{U}$ is randomly replaced with the gate $\mathcal{F}_{i}$, with a sampling overhead of $\kappa^{2}$, where $\kappa^{2} := \sum_{i} \vert a_i \vert$.
For a gate $U$, the smallest possible  $\kappa$ for a given setting $S \in \{\text{LO},\text{LO}\overrightarrow{\text{CC}},\text{LOCC}\}$ is denoted by $\gamma_{S}(U)$.
The sampling overhead is an important quantity to understand in quasiprobability simulation since the overall number of shots, or circuit evaluation, must be scaled by this value to obtain the same amount of error one would expect to get by executing the original circuit.
For $n$ non-local gates, $U_{1},\ldots,U_{n}$, the total sampling overhead is given by $\prod^{n}_{i=1} \gamma_{S}(U_{i})^{2}$.
It should be noted that there are cases where classical communication offers an advantage for multiple cuts of controlled rotation and two-qubit rotation gates.
Additionally, the settings $S$ have the following sampling overhead relationship
\begin{equation}
    \gamma_{\text{LOCC}}(U) \le \gamma_{\text{LO}\overrightarrow{\text{CC}}}(U) \le \gamma_{\text{LO}}(U).
\end{equation}
Despite this known relationship, in this study, we only consider LO since it is the only setting currently implemented in the Qiskit Circuit Cutting addon~\cite{qiskit-addon-cutting}.

\section{Computational Details}
Since we are using the Qiskit Circuit Cutting addon to perform our circuit cutting operations, we chose to use two different packages, Tequila~\cite{kottmann2021tequila} and ffsim~\cite{ffsim}, to generate our electronic structure circuits since the circuits generated by both packages can be directly compiled into Qiskit~\cite{qiskit2024} native circuits.
For the methods implemented in Tequila, which include the UpCCGSD, UpCCD, UCCSD, and SPA+GS ans\"{a}tze, we choose to initialize all model parameters with zeroes.
While these methods could be initialized using HF or M{\o}ller-Plesset second-order perturbation theory (MP2), the choice of circuit parameters does not affect our sampling overhead analysis since the non-local gate cuts in these methods do not correspond to parameterized gates.
Additionally, for the circuits implemented in Tequila, we also examine the effects of cutting circuits using the Jordan-Wigner (JW)~\cite{jordan1928paulische}  and Bravyi-Kitaev (BK)~\cite{bravyi2002fermionic} Fermionic mappings.
For the LUCJ ansatz, we only examine the JW mapping since it is the only one implemented in ffsim, and initialize our circuits using HF, the spin-balanced unitary CJ operator, and the circuit parameters were initialized using CCSD $t_{1}$- and $t_{2}$-amplitudes.
Two additional parameters were explored for the LUCJ ansatz, the number of circuit layers ($L$ in Eq. \ref{eq:LUCJ}) and the circuit layout.
While several choices for circuit layout exist, we choose to examine an all-to-all connectivity, where all possible interaction terms are included, and a heavy-hex topology, since this topology maps natively to IBM devices.
For further information regarding the interaction terms and the effects of various device topologies on the LUCJ ansatz, refer Motta \textit{et al.}~\cite{motta2023bridging}.
Lastly, the Tequila circuits are converted into Qiskit circuits, and any additional classical bits introduced during the conversion process are removed.
The circuits generated in ffsim are already Qiskit circuits but require a decomposition pass using the \textit{reps} parameter set to 1 to interface with the circuit cutting interface.

After choosing the ans\"{a}tze and circuit cutting package, the next choice was determining the molecular systems and basis set to use in our study.
The molecular systems we choose to investigate are hydrogen chains, denoted as H$_{2n}$-chains, for $n=1, 5, 9, 13, 17, 21, 25$, with active spaces corresponding to $2n$ electrons in $2n$ spatial-orbitals, denoted as $(2n, 2n)$. 
An example active space for $n=1$ which corresponds to diatomic hydrogen, H$_{2}$, is highlighted in Fig. \ref{fig:H2example}, where the active space would correspond to two electrons in two spatial orbitals, denoted as (2,2).
We chose to examine these systems since, for each of the ans\"{a}tze, the number of qubits scales linearly with the number of spin-orbitals in the active space, and due to this scaling, we choose to use a minimal basis, STO-3G.
Lastly, all data from this study are available in the following GitHub repository: \href{https://github.com/MSRG/Distributed_Electronic_Structure}{https://github.com/MSRG/Distributed{\textunderscore}Electronic{\textunderscore}Structure}.

\begin{figure}[!t]
    \centering
    \begin{subfigure}{\linewidth}
        \centering
        \includegraphics[width=\linewidth]{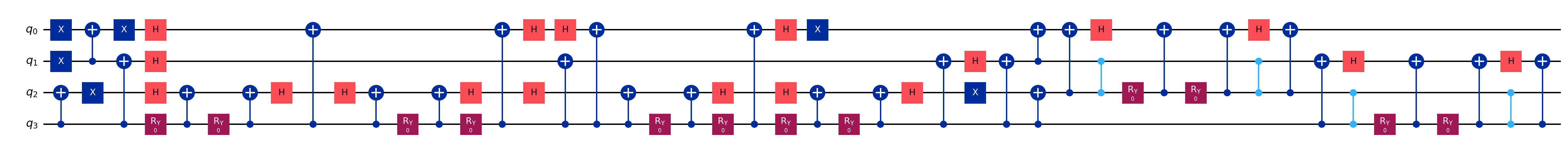}
        \caption{}
        \label{fig:UCCSD_H2_jordan-wigner}
    \end{subfigure}%
    \hfill
    \begin{subfigure}{\linewidth}
        \centering
        \includegraphics[width=\linewidth]{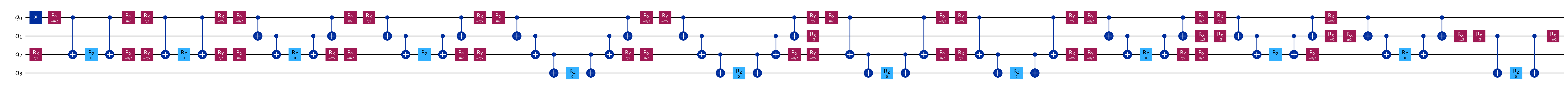}
        \caption{}
        \label{fig:UCCSD_H2_bravyi-kitaev}
    \end{subfigure}
    \hfill
    \begin{subfigure}{\linewidth}
        \centering
        \includegraphics[width=\linewidth]{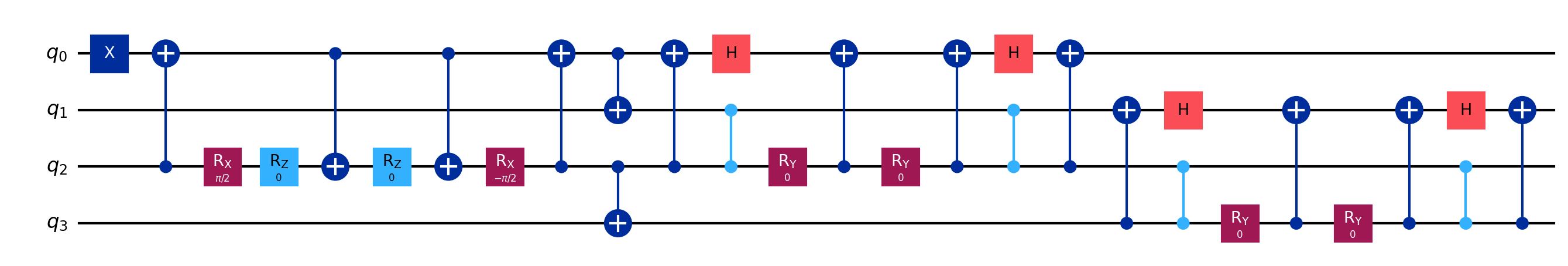}

        \caption{}
        \label{fig:UpCCGSDH2_jordan-wigner}
    \end{subfigure}%
    \hfill
    \begin{subfigure}{\linewidth}
        \centering
        \includegraphics[width=\linewidth]{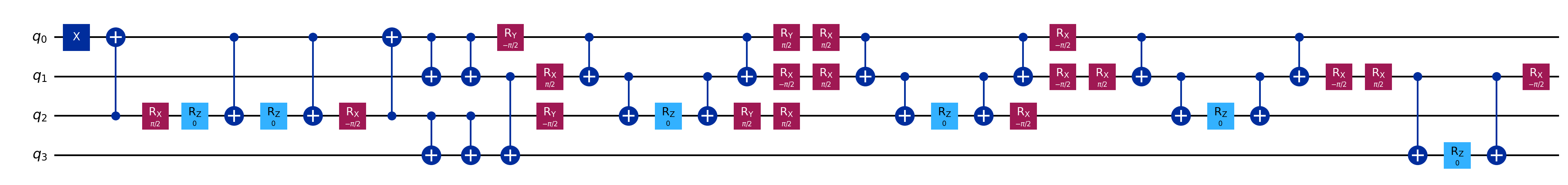}
        \caption{}
        \label{fig:UpCCGSDH2_bravyi-kitaev}
    \end{subfigure}    
    \hfill    
    \begin{subfigure}{0.8\linewidth}
        \centering
        \includegraphics[width=\linewidth]{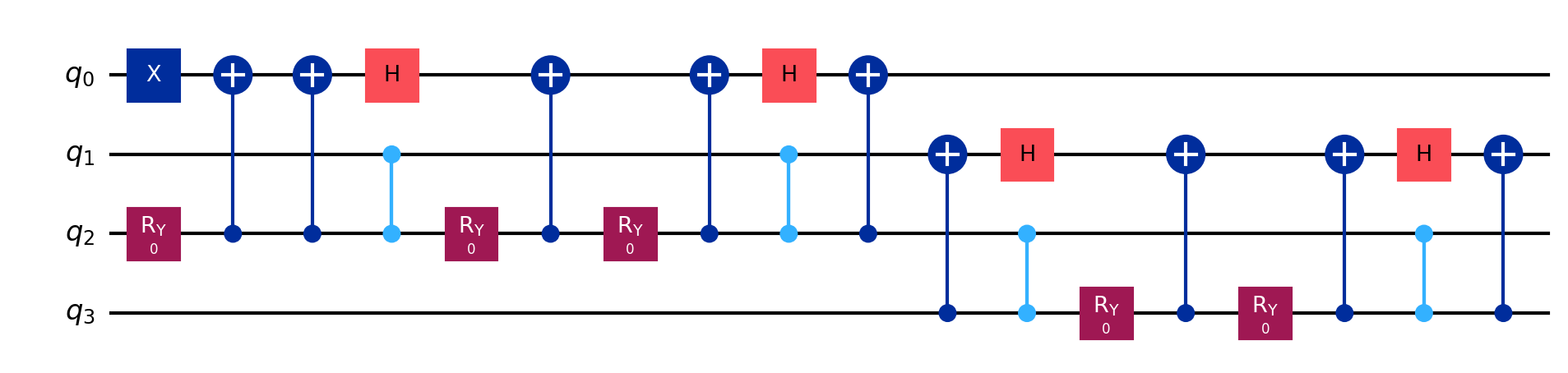}
        \caption{}
        \label{fig:SPA+GSH2_jordan-wigner}
    \end{subfigure}%
    \hfill
    \begin{subfigure}{\linewidth}
        \centering
        \includegraphics[width=\linewidth]{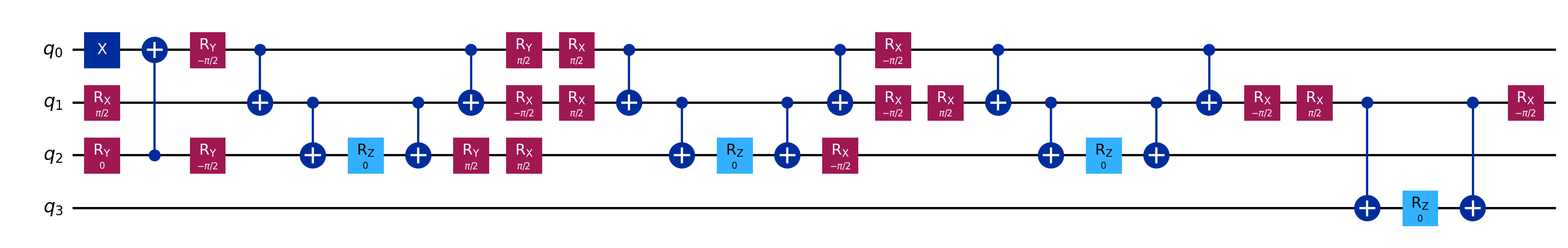}       
        \caption{}
        \label{fig:SPA+GSH2_bravyi-kitaev}
    \end{subfigure}    
    \hfill    
    \begin{subfigure}{0.45\linewidth}
        \centering
        \includegraphics[width=\linewidth]{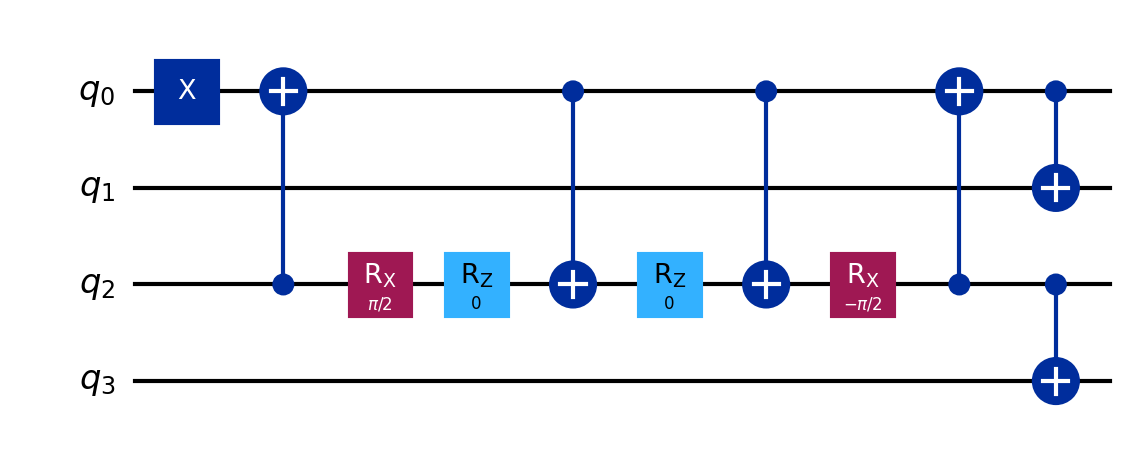}   
        \caption{}
        \label{fig:UpCCDH2_jordan-wigner}
    \end{subfigure}%
    \hfill
    \begin{subfigure}{0.49\linewidth}
        \centering
        \includegraphics[width=\linewidth]{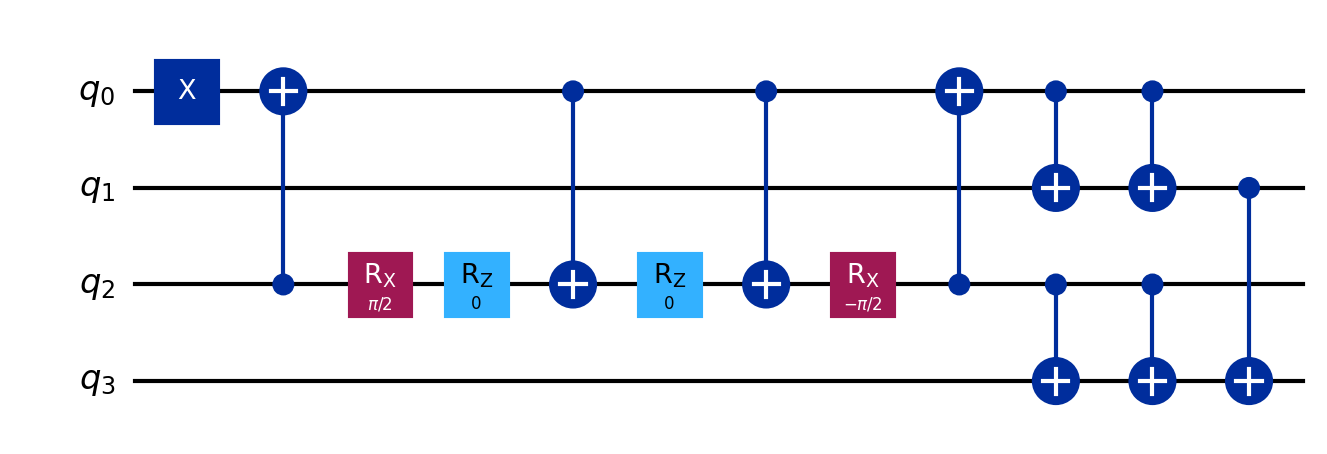}    
        \caption{}
        \label{fig:UpCCDH2_bravyi-kitaev}
    \end{subfigure}
    \hfill
    \begin{subfigure}{0.75\linewidth}
        \centering
        \includegraphics[width=\linewidth]{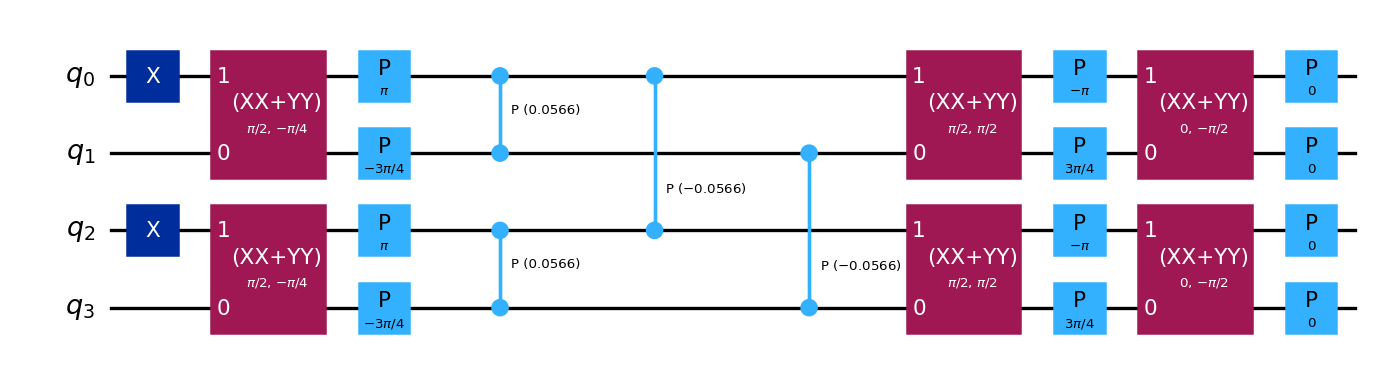}   
        \caption{}
        \label{fig:LUCJ_H2_1_all-to-all}
    \end{subfigure}%
    \hfill
    \begin{subfigure}{0.75\linewidth}
        \centering
        \includegraphics[width=\linewidth]{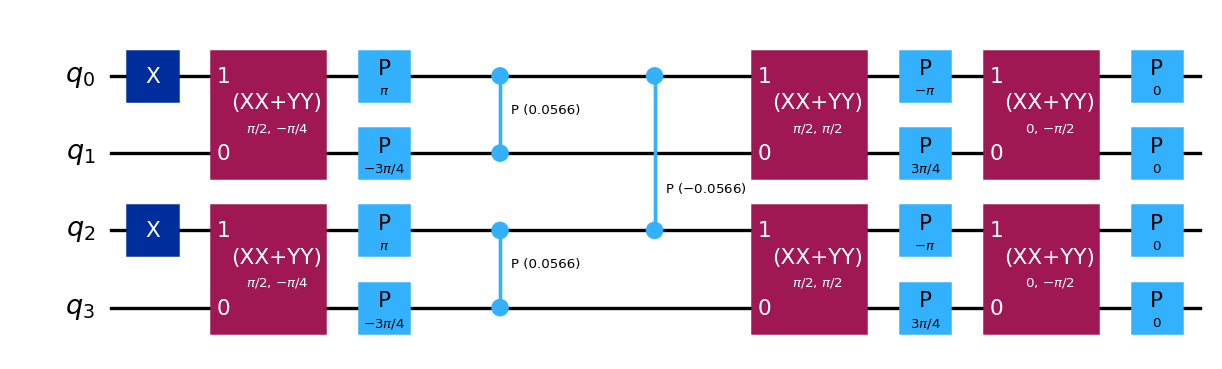}
        \caption{}
        \label{fig:LUCJ_H2_1_heavy-hex}
    \end{subfigure}    
    \caption{Example quantum circuits for H$_{2}$ using a minimal basis, STO-3G, with an active space of two electrons in two orbitals, (2,2). The circuits we explore are as follows: (a) UCCSD/Jordan-Wigner, (b) UCCSD/Bravyi-Kitaev, (c) UpCCD/Jordan-Wigner, (d) UpCCD/Bravyi-Kitaev, (e) UpCCGSD/Jordan-Wigner, (f) UpCCGSD/Bravyi-Kitaev, (g) SPA+GS/Jordan-Wigner, (h) SPA+GS/Bravyi-Kitaev, and (i) LUCJ using an all-to-all and (j) heavy-hex connectivity.}
    \label{fig:H2_layouts}
\end{figure}

\section{Results and Discussion}
In this section, we examine the aptitude of the aforementioned ans\"{a}tze for distribution, which includes insights into the types of gates that are cut, the number of cuts, and the sampling overhead.
We start our analysis using H$_{2}$, where example circuits for each ans\"{a}tz are shown in Fig. \ref{fig:H2_layouts}.
While each of the circuits in Fig. \ref{fig:H2_layouts} is cut in half, between qubits $q_{1}$ and $q_{2}$, the orbitals included in each partition will change, depending on the method.
For the UpCCGSD, UpCCD, UCCSD, and SPA+GS ans\"{a}tze, an order of $\ket{q_{0},q_{1},q_{2},q_{3}} = \ket{\sigma_{\alpha},\sigma_{\beta},\sigma^{\ast}_{\alpha},\sigma^{\ast}_{\beta}} = \ket{1,1,0,0}$, is assumed, where the first subcircuit contains the spin-orbitals associated with the occupied spatial orbital, $\ket{q_{0},q_{1}} = \ket{\sigma_{\alpha},\sigma_{\beta}} = \ket{1,1}$, and the second corresponds to the unoccupied orbitals, $\ket{q_{2},q_{3}} = \ket{\sigma^{\ast}_{\alpha},\sigma^{\ast}_{\beta}} = \ket{0,0}$.
In ffsim, the LUCJ ansatz has a different ordering, where the full circuit takes the following form $\ket{q_{0},q_{1},q_{2},q_{3}} =\ket{\sigma_{\alpha},\sigma^{\ast}_{\alpha},\sigma_{\beta},\sigma^{\ast}_{\beta}}=\ket{1,0,1,0}$. 
At the same time, the first subcircuit includes the alpha spin-orbitals, $\ket{q_{0},q_{1}} =\ket{\sigma_{\alpha},\sigma^{\ast}_{\alpha}}=\ket{1,0}$, and the second subcircuit contains the beta spin-orbitals, $\ket{q_{2},q_{3}} =\ket{\sigma_{\beta},\sigma^{\ast}_{\beta}}=\ket{1,0}$.

For each of the following ans\"{a}tze, UpCCGSD, UpCCD, UCCSD, and SPA+GS, the gate cuts will correspond to controlled-Z (CZ) or controlled-X (CNOT/CX) gates.
For both CZ and CX gates, the sampling overhead factor is $\gamma(\text{CX})^{2}=\gamma(\text{CZ})^{2}=3^{2}=9$, as defined in~\cite{mitarai2021constructing,schmitt2025cutting}, therefore, the overall sampling overhead for each circuit will scale exponentially with the number of nonlocal gate cuts ($N_{\text{cuts}}$) defined as $\prod_{i=1}^{N_{\text{cuts}}}9$ or $9^{N_{\text{cuts}}}$.
Starting with the UCCSD ansatz using the JW encoding (Fig. \ref{fig:UCCSD_H2_jordan-wigner}), this circuit contains 16 controlled-X and 2 controlled-Z gates that are cut, corresponding to a sampling overhead of $9^{16+2}=1.5009\times10^{17}$.
The corresponding circuit using BK encoding (Fig. \ref{fig:UCCSD_H2_bravyi-kitaev}) has a higher sampling overhead of $9^{24}=7.9766\times10^{22}$, due to the circuit containing 24 controlled-X gates that must be cut to provide two partitions. 
The ansatz with the second worst scaling is the UpCCGSD method, where the JW (Fig. \ref{fig:UpCCGSDH2_jordan-wigner}) encoding has 12 controlled-X and 2 controlled-Z cuts with a sampling overhead of $9^{12+2}=2.2877\times10^{13}$, while the BK (Fig. \ref{fig:UpCCGSDH2_bravyi-kitaev}) has a slightly smaller sampling overhead of $9^{13} = 2.5419\times 10^{12}$ due to 13 controlled-X gate cuts.
The next circuit examined is SPA+GS, where the JW (Fig. \ref{fig:SPA+GSH2_jordan-wigner}) circuit contains 11 cuts, 9 of which correspond to controlled-X cuts and 2 of which correspond to controlled-Z cuts, with a sampling overhead of $9^{11}= 3.1381\times10^{10}$, and the BK circuit (Fig. \ref{fig:SPA+GSH2_bravyi-kitaev}) contains 9 controlled-X cuts for a sampling overhead of $9^{9}=3.8742\times 10^{8}$.
Among the circuits that contain controlled-X or controlled-Z gate cuts, the UpCCD ansatz requires the lowest number of gate cuts, where the circuit using the JW (Fig. \ref{fig:UpCCDH2_jordan-wigner}) mapping contains 4 controlled-X gate cuts with a sampling overhead of $10^4=6.5610\times10^{3}$ and the BK (Fig. \ref{fig:UpCCDH2_bravyi-kitaev}) encoding contains 5 controlled-X gate cuts with a sampling overhead of $5.9049\times10^{4}$.
The lowest sampling overhead corresponds to the LUCJ ansatz using an all-to-all (Fig. \ref{fig:LUCJ_H2_1_all-to-all}) and heavy-hex (Fig. \ref{fig:LUCJ_H2_1_heavy-hex}) layouts.
For these circuits, the two-qubit gate cuts correspond to controlled-phase gates (CP), with a sampling overhead of $\gamma(\text{CP}(\theta))^2=(1 + 2 \vert \sin{(\theta/2)} \vert )^{2}$, as defined in~\cite{schmitt2025cutting}, where the total sampling overhead of the circuit can be defined as $((1 + 2 \vert \sin{(\theta/2)} \vert )^{2})^{N_{\text{cuts}}}$.
The sampling overhead of the all-to-all example, with both CP gates being cut having $\theta=-0.0566$, to overall sampling overhead for this circuit is $((1 + 2 \vert \sin{(-0.0566/2)} \vert )^{2})^{2}=1.2463$.
The heavy-hex layout contains one CP gate, with an angle of $\theta=-0.0566$, that is cut for a total sampling overhead of $(1 + 2 \vert \sin{(-0.0566/2)} \vert )^{2}=1.1164$.

The sampling overheads can provide insights into what types of circuits can be reasonably implemented using LO, since the number of circuit evaluations (shots) must scale with the sampling overheads. 
For example, consider the case of UCCSD using the JW encoding, which has a sampling overhead of $1.5009\times10^{17}$, where circuits are often evaluated using $10^{3}-10^4$ shots to obtain reliable results from noisy quantum devices.
Even for this minimal case, this would require $(10^{3}-10^4) \cdot (1.5009\times10^{17})$ circuit evaluations to obtain the same level of accuracy as the uncut circuit using $10^{3}-10^4$ shots.
Alternatively, the LUCJ ansatz, partitioned using LO, could be implemented on the current generation of devices due to small sampling overheads.
For example, the partitioned circuit with all-to-all connectivity would require $1.2463 \times 10^{3}-10^4$ evaluations to obtain results similar to the unpartitioned circuit using $10^{3}-10^4$ shots.
The large sampling overheads for UpCCGSD, UpCCD, UCCSD, and SPA+GS highlight that even when using a minimal active and basis set for a small molecule, H$_{2}$, the unpartitioned quantum circuit is more efficient to implement than the LO case.

\begin{figure}[!t]
    \centering
    \begin{subfigure}{\linewidth}
        \centering
        \includegraphics[width=\linewidth]{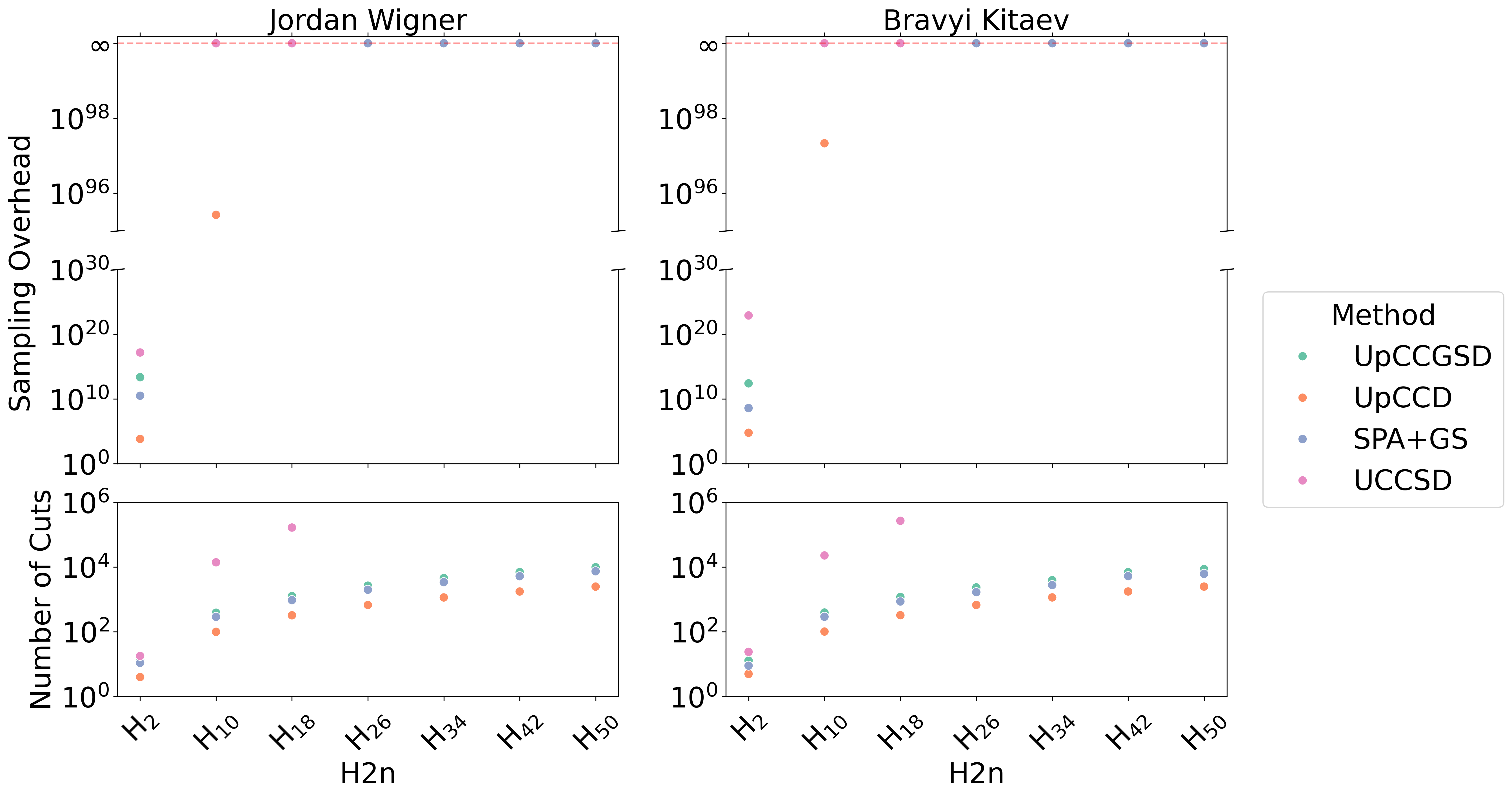}
        \caption{}
        \label{fig:H2n_cuts_overhead}
    \end{subfigure}%
    \hfill    
    \begin{subfigure}{\linewidth}
        \centering
        \includegraphics[width=0.75\linewidth]{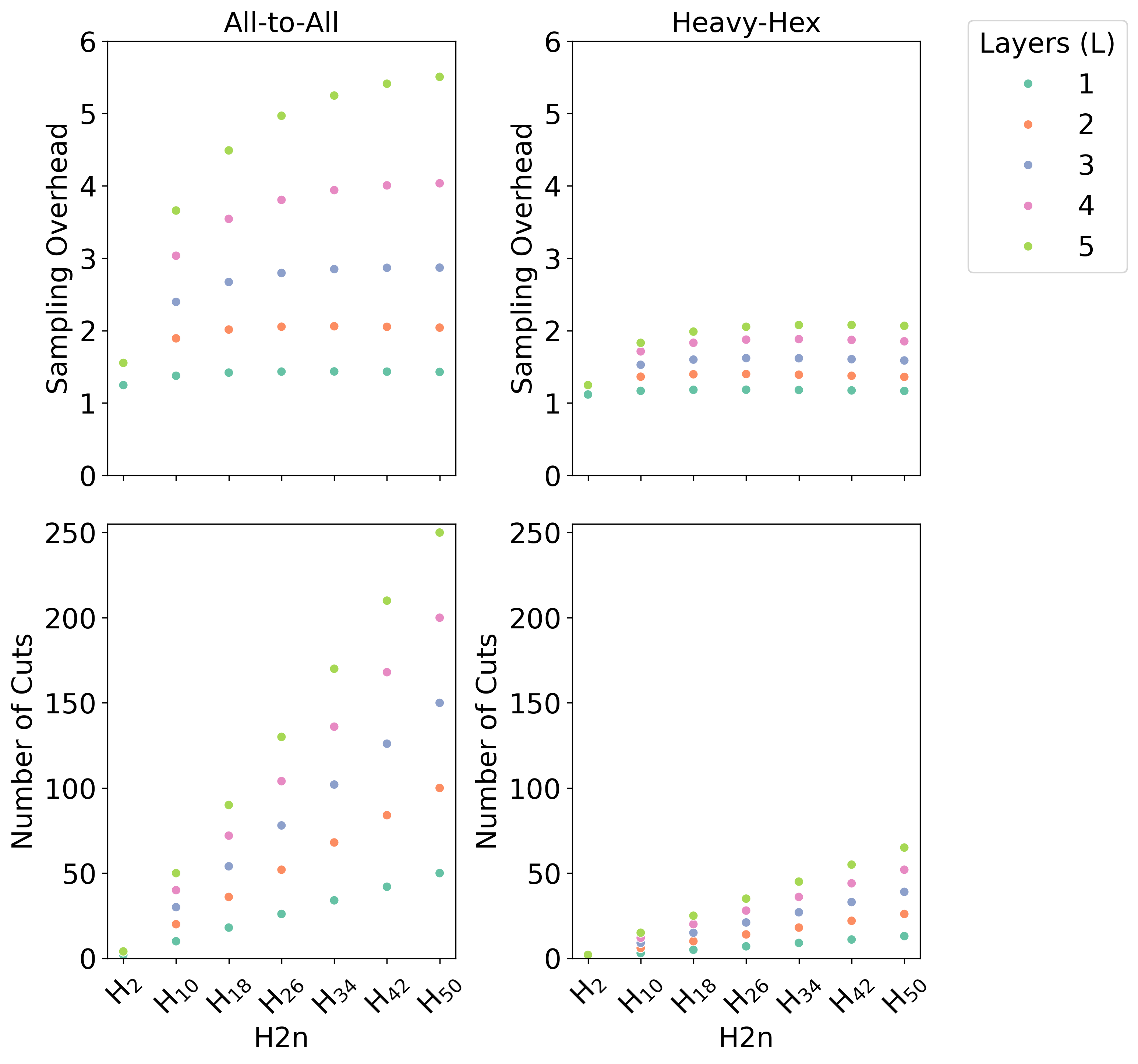}
        \caption{}
        \label{fig:LUCJ_H2n_cuts_overhead}
    \end{subfigure}%
    \caption{(a) The sampling overhead (top row) and number of cuts (bottom row) for various methods implemented in Tequila using both the Jordan-Wigner and Bravyi-Kitaev Fermionic encodings. (b) The sampling overhead (top row) and number of cuts (bottom row) for the local unitary cluster Jastrow (LUCJ) ansatz with all-to-all (left column) and heavy-hex (right column) using circuit repetitions of 1-5 for various hydrogen chains.}
    \label{fig:H2n_cuts_overhead_all}
\end{figure}

Using the insights provided by H$_{2}$, we will show that for the remaining H$_{2n}$-chains, the UpCCGSD, UpCCD, UCCSD, and SPA+GS ans\"{a}tze are unfeasible for distribution using local operations, whereas LUCJ offers a method with favorable scaling regarding system size.
Across all systems and methods examined, the minimum number of cuts corresponds to 1 for H$_{2}$ using the LUCJ ansatz with the heavy-hex layout, the mean is 4904.55 cuts with a standard deviation of 29262.70, and the maximum number of cuts corresponds to UCCSD using the BK mapping on H$_{18}$. 
Note that, we use all H$_{2n}$ chains previously mentioned with all ans\"{a}tze, except for the UCCSD method, where the maximum system size analyzed was H$_{18}$, due to the scaling of the circuit depth with system size.
While the number of cuts scales as expected for each ansatz, i.e., H$_{2}$ will require the fewest cuts and H$_{50}$ will require the most cuts, the convergence of the sampling overheads varies by method.
This is highlighted in Fig. \ref{fig:H2n_cuts_overhead}, where, for both encoding methods, the only molecule that does not contain an infinite sampling overhead is H$_{2}$.
In this case, any sampling overhead greater than $1.7977 \times 10^{308}$ is considered infinite since this is the maximum value set for double precision floats in NumPy~\cite{harris2020array}.
Across all systems examined, with active spaces ranging from (2,2) to (50,50), the LUCJ ansatz offers favorable scaling regarding the sampling overhead (Fig. \ref{fig:LUCJ_H2n_cuts_overhead} top row) and number of cuts (Fig. \ref{fig:LUCJ_H2n_cuts_overhead} bottom row) for both the all-to-all (left column) and heavy-hex (right column) layouts.
The maximum sampling overhead corresponds to 5.5059 for H$_{50}$ with 5 layers for the all-to-all connectivity, while the analogous heavy-hex case has a sampling overhead of 2.0776.

Of all the circuits analyzed, it is clear that, regardless of the device topology, LUCJ offers a favorable method for distribution based on the sampling overheads.
Additionally, methods, such as UCCSD, that require cutting CX or CZ gates offer unfavorable methods for distribution due to the exponential scaling of the sampling overhead with the number of gates cut.
To avoid the unfavorable scaling of these methods that contain CX or CZ gate cuts, new methods must be developed.
To this end, recent work from Xue \textit{et al.}, implemented a distributed unitary selective coupled cluster (dUSCC) using pseudo-commutativity of Trotterization for distribution among multiple QPUs~\cite{xue2025efficient}.
This work, along with the insights provided in this study, highlight the need for developing new methods for parallelizing quantum algorithms for electronic structure theory calculations.

\section{Outlook and Conclusions}
In this study, we highlighted the difficulty of implementing LO on existing implementations of quantum algorithms for electronic structure theory.
We showed that for the smallest diatomic molecule, H$_{2}$, using a minimal active space and basis set, certain algorithms, such as UCCSD, UpCCGSD, and SPA+GS, require an unreasonable amount of circuit evaluations when compared to unpartitioned circuits.
Despite SPA+GS and LUCJ offering promising scaling for H$_{2}$, only the LUCJ algorithm scales reasonably with system size, as SPA+GS converges towards infinity.
Due to the favorable scaling of the LUCJ ansatz, we are currently exploring distribution using LO within a quantum-centric supercomputing framework~\cite{robledo-moreno_chemistry_2024}.



\begingroup
\small  
\bibliographystyle{IEEEtran}
\bibliography{quantum_week}
\endgroup

\end{document}